\documentclass{article}

\usepackage[dblblindworkshop, final]{neurips_2025}
\workshoptitle{Machine Learning and the Physical Sciences}

\usepackage[utf8]{inputenc} 
\usepackage[T1]{fontenc}    
\usepackage{hyperref}       
\usepackage{url}            
\usepackage{booktabs}       
\usepackage{amsfonts}       
\usepackage{nicefrac}       
\usepackage{microtype}      
\usepackage{xcolor}         

\usepackage{amsmath}
\usepackage{graphicx}
\usepackage{subcaption}

\usepackage{tikz}
\usetikzlibrary{arrows.meta,positioning,fit}
\usepackage{rotating}
\usepackage{tikz}
\usetikzlibrary{positioning,calc,arrows.meta}
\usepackage{subcaption}   
\usepackage{graphicx}     
\usetikzlibrary{fit}
\usetikzlibrary{positioning,calc,arrows.meta,fit,decorations.pathreplacing}

\title{Transfer Learning Beyond the Standard Model}

\author{%
\textbf{Veena Krishnaraj}$^1$ \quad\textbf{Adrian E. Bayer}$^{2,1}$  \quad \textbf{Christian Kragh Jespersen}$^{1}$ \quad \textbf{Peter Melchior}$^{1}$ \\
$^1$Princeton University, Princeton, NJ 08544, USA \\
$^2$Flatiron Institute, New York, NY 10010, USA \\
\texttt{\{vk9342,abayer,ckragh,peter.melchior\}@princeton.edu}
}

\begin{document}

\maketitle

\begin{abstract}
Machine learning enables powerful cosmological inference but typically requires many high-fidelity simulations covering many cosmological models. Transfer learning offers a way to reduce the simulation cost by reusing knowledge across models. We show that pre-training on the standard model of cosmology, $\Lambda$CDM, and fine-tuning on various beyond-$\Lambda$CDM scenarios---including massive neutrinos, modified gravity, and primordial non-Gaussianities---can enable inference with significantly fewer beyond-$\Lambda$CDM simulations. However, we also show that negative transfer can occur when strong physical degeneracies exist between $\Lambda$CDM and beyond-$\Lambda$CDM parameters. We consider various transfer architectures, finding that including bottleneck structures provides the best performance.
Our findings illustrate the opportunities and pitfalls of foundation-model approaches in physics: pre-training can accelerate inference, but may also hinder learning new physics.
\end{abstract}

\section{Introduction}

Simulation-based inference (SBI) has been successfully adopted in cosmology to infer the standard model ($\Lambda$CDM) parameters from large-scale structure surveys \citep{Hahn:2023udg}.
A key goal of Stage-IV surveys is to detect physics beyond the standard model---such as massive neutrinos, modified gravity, and primordial non-Gaussianities \citep{desicollaboration2024desi2024viicosmological}. Accurately testing beyond-$\Lambda$CDM extensions requires large suites of computationally expensive simulations, often far more expensive than their $\Lambda$CDM counterparts, creating a major bottleneck.

A promising way to alleviate this challenge is transfer learning, where knowledge acquired in one domain is reused to accelerate learning in another \citep{transfer_learing_overview}. In cosmology, transfer learning has recently been applied between low and high fidelity simulations of the same underlying physics \citep{cosm_transfer_learning_ex, Hikida:2025xuw, Thiele:2025mlo}. 
In this work, we ask a more ambitious question: \textbf{can transfer learning enable machine learning models to generalize to new physics?} Specifically, we investigate fine-tuning neural networks trained on $\Lambda$CDM to perform parameter inference beyond $\Lambda$CDM. 
In a sense, our study probes whether $\Lambda$CDM can serve as a foundation model upon which physics beyond the standard model can be fine-tuned. 

Despite its promise, transfer learning has been shown to sometimes hinder performance in a phenomenon known as \textit{negative transfer} \citep{negative_transfer_overview}. Whether transfer succeeds depends on the relationship between the source and target domains: if the target involves genuinely new physics not represented in the pre-trained model, or if strong parameter degeneracies obscure the relevant signals, transfer can fail or mislead. We thus explore different transfer architectures, including bottleneck or “dummy” units, to balance reuse of $\Lambda$CDM features with the flexibility to capture new physics.

\section{Methods}
\label{sec:methods}

\begin{figure}[t]
\centering
\includegraphics[width=0.8\columnwidth]{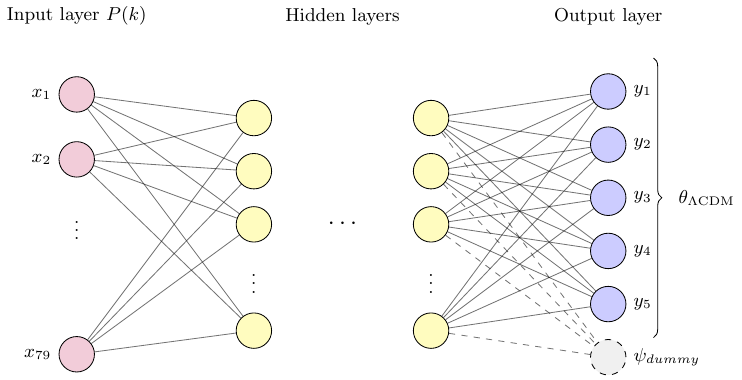}
\caption{Dummy network architecture. The model takes the (marked) power spectrum $P(k)$ as input and outputs cosmological parameters $\theta_{\Lambda{\rm CDM}}$. Additional latent ``dummy'' nodes $\psi_{\rm dummy}$ are included in the output layer to provide extra representational capacity for fine-tuning.}
\label{fig:nn_diagram}
\end{figure}

We consider three different beyond-$\Lambda$CDM (fine-tuning) examples: massive neutrinos ($M_\nu$), modified gravity ($f(R)$), and primordial non-Gaussianities ($f_{\rm NL}^{\rm equilateral}$ and $f_{\rm NL}^{\rm local}$). 
We use the Quijote simulations \citep{Quijote_sims}. For the $\Lambda$CDM (pre-training) simulations, we vary 5 cosmological parameters: $\Omega_m \in [0.10,\ 0.50],\ \Omega_b \in [0.02,\ 0.08],\ h \in [0.50,\ 0.90],\ n_s \in [0.80,\ 1.20], \ \sigma_8 \in [0.60,\ 1.00]$, and fix $M_{\nu} = 0{\rm eV}$, $w = -1$, $f_{R0}=0$, and $f_{\rm NL}=0$. 
For each beyond-$\Lambda$CDM (fine-tuning) example, a separate Latin Hypercube of simulations is used where both the $\Lambda$CDM and beyond-$\Lambda$CDM parameters are varied (except in the case of local-$f_{\rm NL}$, where only $f_{\rm NL}$ is varied, to assess the impact of a mismatch in the distribution of $\Lambda$CDM parameters during transfer).
We provide a thorough description of the simulation setup in Appendix \ref{app:data}.

We use a fully connected neural network to predict cosmological parameters from the matter power spectrum (or marked power spectrum \citep{Massara:2020pli}). The input to the network is a vector of 79 bins linearly spaced in the range $k \in [0.0089, 0.5]\,h/{\rm Mpc}$. 
All target parameters are linearly normalized to the range [0, 1]. 
The simulations for a given cosmology are divided into training, validation, and testing datasets, comprised of 70\%, 15\%, and 15\% of the total dataset respectively. 
We further subsample the training set to investigate the performance as a function of the number of pre-training and fine-tuning simulations, while the validation and test sets remain fixed.

Our model consists of a fully connected neural network with up to three hidden layers, each consisting of a LeakyReLU activation (slope 0.2). A sigmoid activation function is applied to the output layer to match the $[0, 1]$ normalized targets. Training minimizes mean squared error (MSE) using the AdamW optimizer ($\beta_1 = 0.5$, $\beta_2 = 0.999$), with batch size 32 and early stopping if validation loss does not improve by more than $10^{-6}$ after 50 epochs, with a maximum limit of 1000 epochs. We use Optuna~\citep{optuna_2019} to tune the number of layers (up to 3), neurons per hidden layer (4-500), learning rate, weight decay, and dropout, running 100 trials with TPE sampling and pruning. 

We implement a two-stage transfer learning approach. First we train the network on the $\Lambda$CDM simulation set. Crucially, we include dummy nodes $\psi_{dummy}$ in the pre-training network to output the same number of parameters as the corresponding beyond-$\Lambda$CDM model. For pre-training, the MSE is computed only using the $\Lambda$CDM parameters, thus the extra nodes are dummies. 
During pre-training we allow for learning rates in $[10^{-5},10^{-1}]$. 
In the second stage we fine-tune the network on the beyond-$\Lambda$CDM dataset with initialized weights from the pre-trained network, and using the dummy nodes for the beyond-$\Lambda$CDM parameters, reducing the learning rate range to $[10^{-6},10^{-3}]$. 
Fig \ref{fig:nn_diagram} depicts our network setup. 

Our choice of including dummy nodes in the pre-training is motivated by prior work in representation learning, where additional latent units, or bottleneck structures, can improve transferability and mitigate negative transfer \citep{yosinski2014transferablefeaturesdeepneural, bengio2014representationlearningreviewnew,ho2023informationorderedbottlenecksadaptivesemantic}, and is conceptually related to the modular “head” architectures used in foundation models that enable flexible adaptation to diverse downstream tasks \citep{devlin2019bertpretrainingdeepbidirectional, radford2021learningtransferablevisualmodels}.
We also investigated two other typical pre-training architectures which we found to be suboptimal: one without any dummy nodes, and one where we fix the pre-trained weights and attach a trainable inference head instead. Further details are provided in Appendix \ref{app:arch}.

\section{Results}
\label{sec:results}

\begin{figure}[t] 
    \centering
    \begin{subfigure}{0.99\linewidth}
        \centering
        \includegraphics[width=\linewidth]{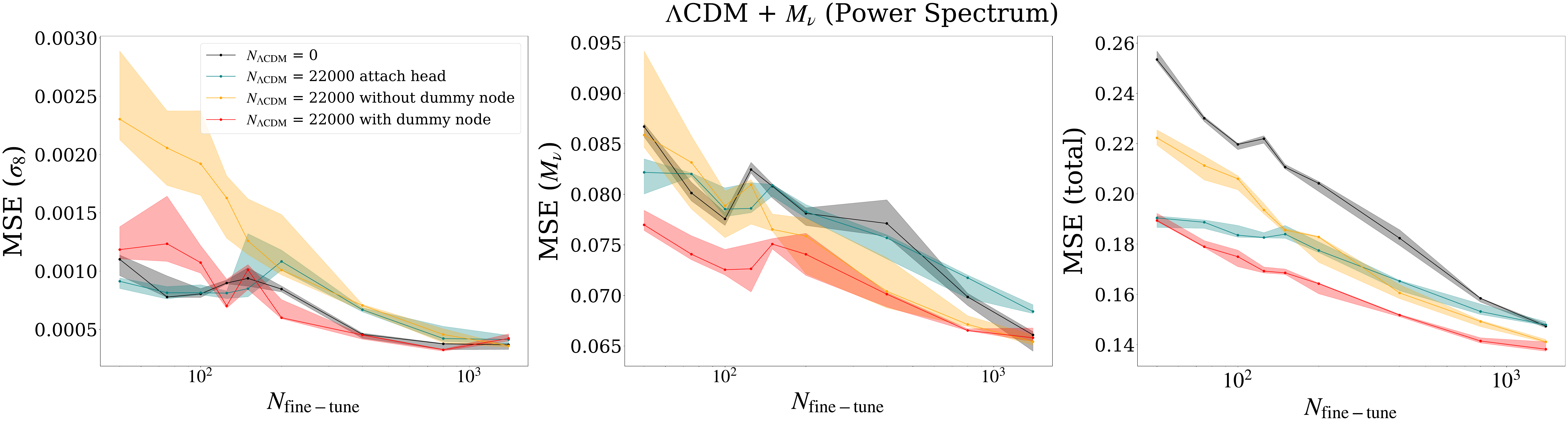}
        \label{fig:mse_nwLH_Pk}
    \end{subfigure}
    \hspace{0.0\linewidth}
    \begin{subfigure}{0.99\linewidth}
        \centering
        \includegraphics[width=\linewidth]{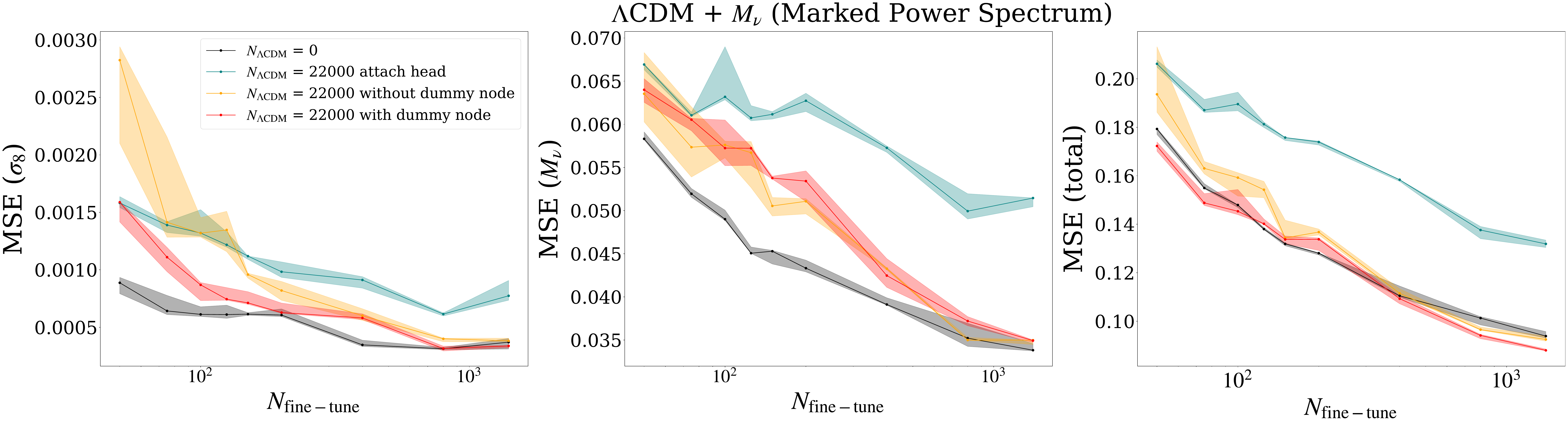}
        \label{fig:mse_nwLH_MPk}
    \end{subfigure}
    \hspace{0.0\linewidth}
    \vspace{-2em}
    \caption{Test MSE as a function of the number of fine-tuning simulations for the massive neutrino cosmology using standard (top) and marked (bottom) power spectra for $\sigma_8$ (left), $M_\nu$ (center), and the total MSE across all normalized parameters (right).
    Transfer learning using a dummy node (red) always outperforms the result with no transfer learning (black) in terms of the total MSE, however negative transfer occurs for the marked power spectrum for $\sigma_8$ and $M_\nu$ due to the physical degeneracy between $M_\nu$ and $\sigma_8$.
    Other transfer learning architectures (teal, yellow) are suboptimal and result in more severe negative transfer.}
    \vspace{-0.8em}
    \label{fig:mse_nwLH}
\end{figure}

Fig~\ref{fig:mse_nwLH} shows the test MSE as a function of the number of beyond-$\Lambda$CDM simulations used to train the fine-tuning network for the $M_\nu$ extension to $\Lambda$CDM.  
Each point represents the median MSE of the fine-tuning network’s top 10 performing models, with error bars indicating the 16th and 84th percentiles.
$22{,}000$ simulations are used for pre-training.
We consider the MSE on two individual parameters, $\sigma_8$ and $M_\nu$, as well as the total MSE across all parameters.
Transfer learning using a dummy node (red) always outperforms no transfer learning (black) in terms of the total MSE, with almost an order of magnitude less simulations required to achieve a given total MSE in the case of the power spectrum.
However, in the case of the marked power, the MSE on $\sigma_8$ and $M_\nu$ is worse when performing transfer learning. This \textit{negative transfer} occurs because of the  physical degeneracy between $\sigma_8$ and $M_\nu$ \citep{Bayer:2021iyb}---the pre-trained network has learned what features in the data to associate with $\sigma_8$ in the absence of neutrino mass, and then has to unlearn some of these features and associate them with $M_
\nu$ upon fine-tuning. This occurs for the marked power which is very sensitive to $\sigma_8$ and $M_\nu$ \citep{Massara:2020pli}, whereas the power spectrum alone is less sensitive \citep{Bayer:2021iyb} and thus the introduction of $M_\nu>0$ does not confuse the pre-trained network. We explicitly show this negative learning in by performing a feature analysis in  Appendix~\ref{app:shap}.
Other transfer learning architectures---without the dummy node, or by attaching a head---perform worse in the limit of large number of simulations, and can cause an even larger negative transfer: in particular, head attachment suffers from negative transfer in terms of the total MSE, as the frozen weights of the pre-trained network enforce a representation which is too rigidly aligned with $\Lambda$CDM and thus the tuneable head is unable to transfer beyond it. However, for very few simulations ($<10^2$) head attachment is of comparable quality to the other methods.

\begin{figure}[t!]
    \centering
\begin{subfigure}{0.32\linewidth}
        \centering
        \includegraphics[width=\linewidth]{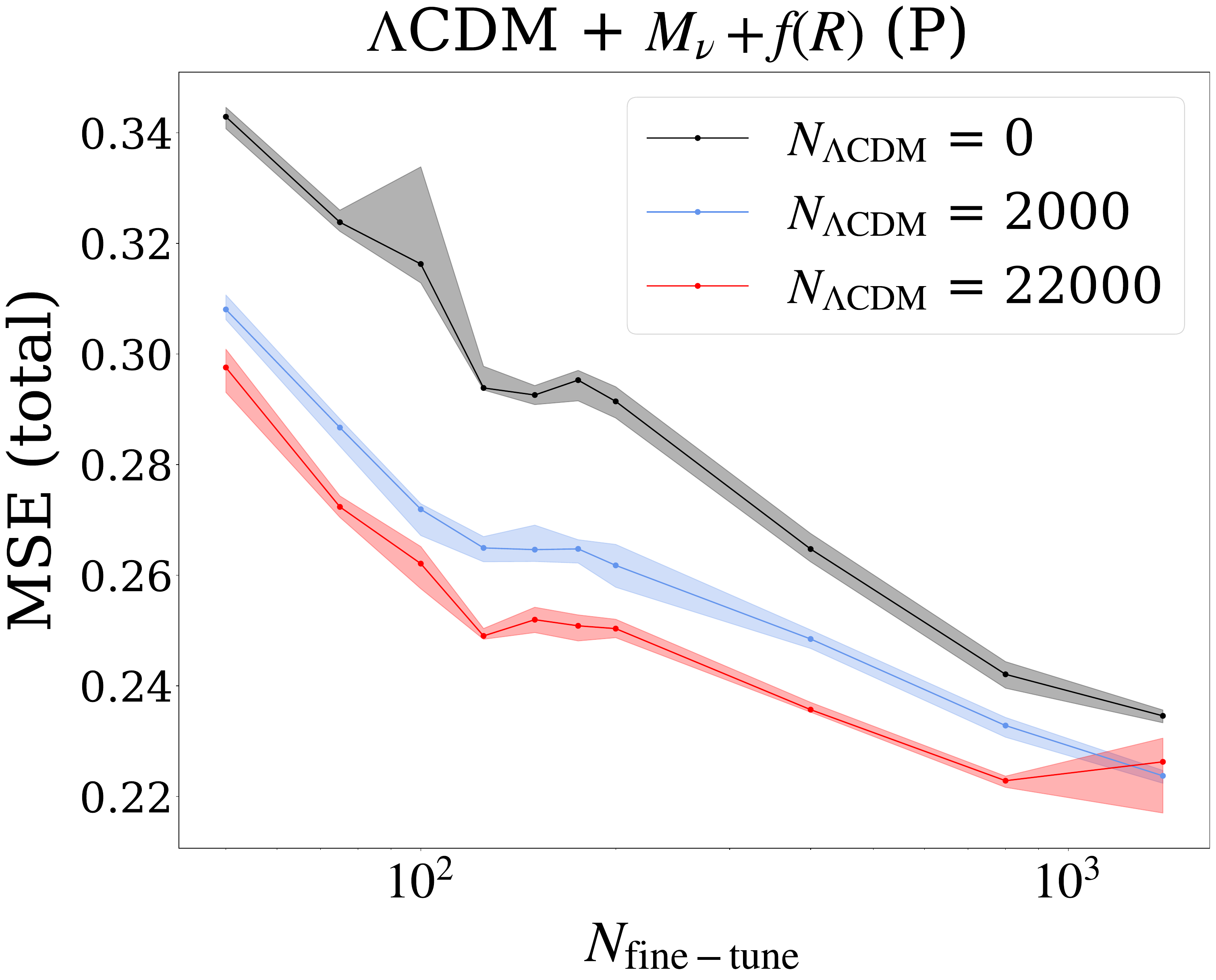}
        \caption{}
        \label{fig:mse_fR_Pk}
    \end{subfigure}
    \hspace{0.0\linewidth}
    \begin{subfigure}{0.32\linewidth}
        \centering
        \includegraphics[width=\linewidth]{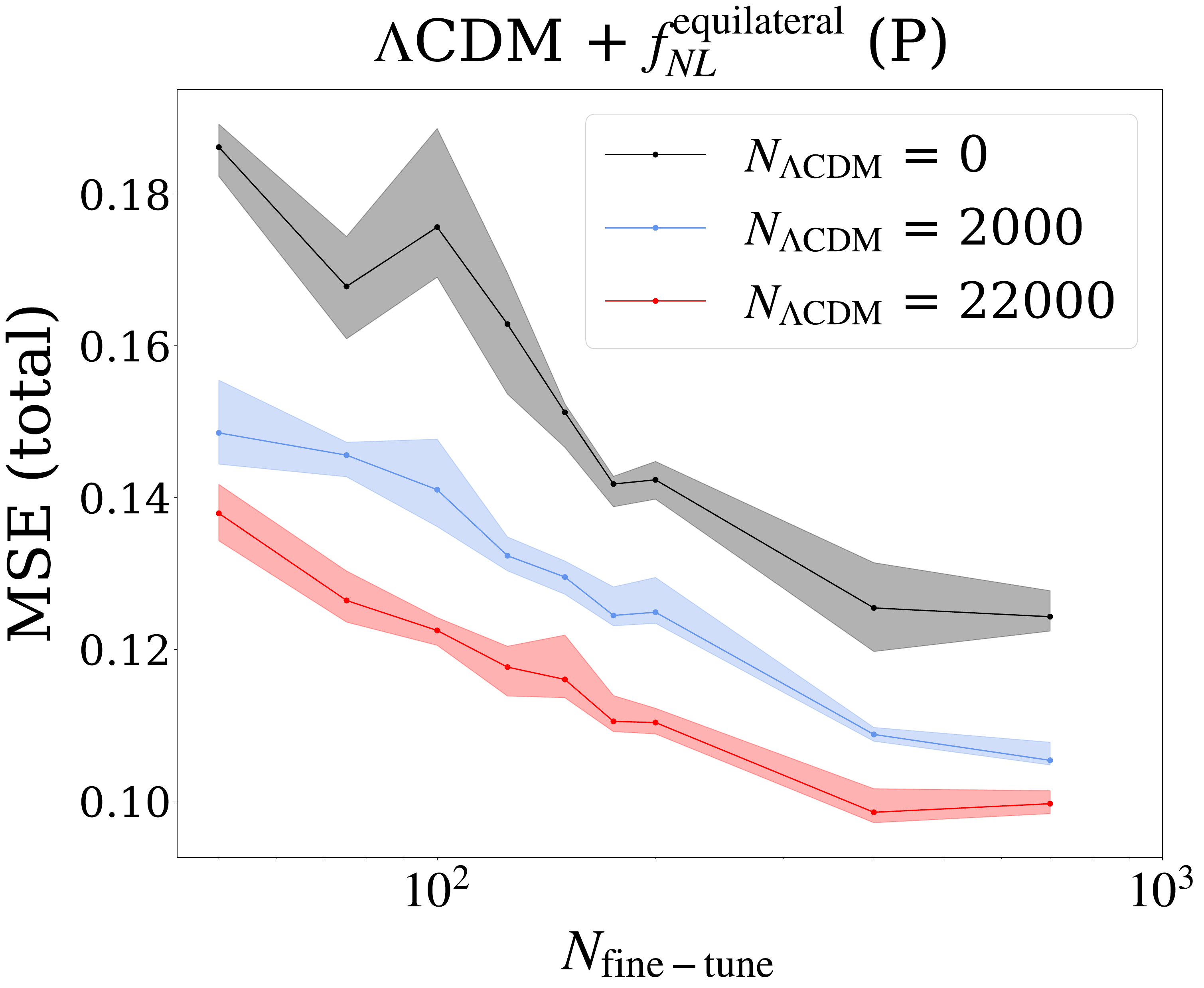}
        \caption{}
        \label{fig:mse_EQ_Pk}
    \end{subfigure}
    \hspace{0.0\linewidth}
    \begin{subfigure}{0.32\linewidth}
        \centering
        \includegraphics[width=\linewidth]{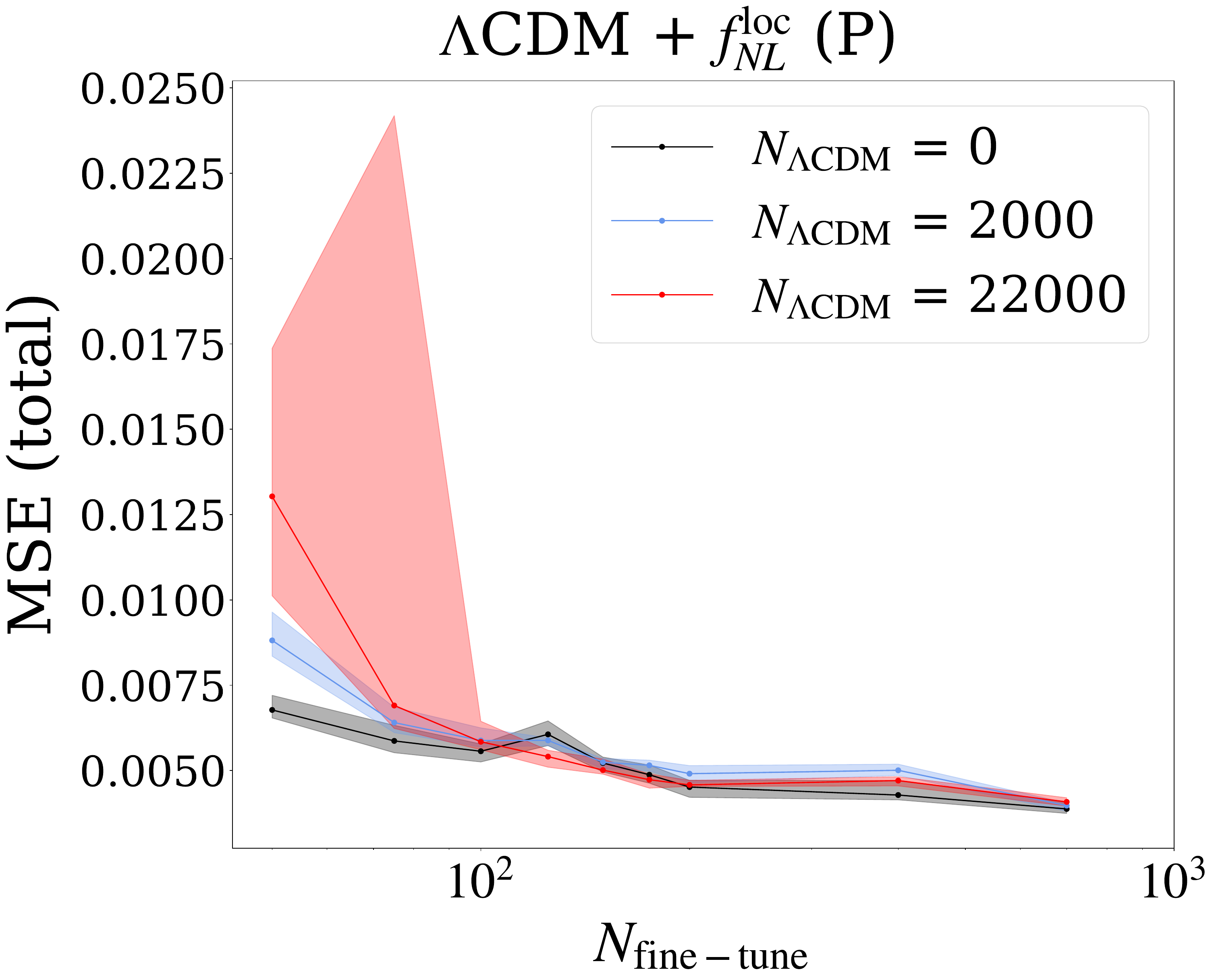}
        \caption{}
        \label{fig:mse_LC_Pk}
    \end{subfigure}
    \caption{Total MSE across all normalized parameters for modified gravity (left), equilateral (center), and local (right) primordial non-Gaussianity cosmologies. The colored lines represent different pre-training set sizes, which outperform the model trained directly on beyond-$\Lambda$CDM without transfer learning (black), except in the case of local $f_{\rm NL}$ due to the prior.}
    \label{fig:mse_beyond_lcdm}
\end{figure}

Having determined the dummy node approach to be best, we analyze the other beyond-$\Lambda$CDM scenarios with this method, exploring the effect of the number of pre-training simulations. Fig \ref{fig:mse_beyond_lcdm} shows the total MSE for the remaining beyond-$\Lambda$CDM cosmologies.
In the modified gravity case, results are similar to the massive neutrino case, with significant gains. There is also an increase in performance in the equilateral $f_{\rm NL}$ case, where degeneracies are mild. In the local non-Gaussianity case, the Quijote simulations do not vary the $\Lambda$CDM parameters, and only vary $f_{\rm NL}^{\rm local}$, thus transfer learning has little advantage---while this result is simply due to the simulation prior, we include the result to show the effect of different priors on the $\Lambda$CDM parameters between the pre-training and fine-tuning simulations. In all examples we find that even 2{,}000 pre-training simulations is enough to see benefits from transfer learning, with further improvements when using 22{,}000.

Full per-parameter MSE results are provided in Appendix~\ref{app:additional_results}, Figs~\ref{fig:mse_nwLH_full}–\ref{fig:mse_nwLH_ft_test_full}.

\section{Discussion and Conclusions}



In this work, we investigated the effectiveness of transfer learning for cosmological parameter inference beyond the standard $\Lambda$CDM model. Using a two-stage approach, we pre-trained neural networks on large $\Lambda$CDM simulation datasets and fine-tuned them on much smaller, computationally expensive beyond-$\Lambda$CDM simulations. We considered cosmologies with massive neutrinos, modified gravity, and primordial non-Gaussianities, using both power spectra and marked power spectra.

We find that transfer learning can reduce simulation requirements by up to an order of magnitude, but its success is dependent on the underlying parameter space. In models with fewer degeneracies—such as equilateral-type primordial non-Gaussianity—transfer learning improves inference across most parameters. By contrast, in scenarios with strong degeneracies, such as massive neutrino cosmologies where $\sigma_8$ and $M_\nu$ are entangled, transfer learning can lead to negative transfer, particularly when using a summary which is very sensitive to $\sigma_8$ and $M_\nu$. Among the architectures tested, we found that introducing additional latent units, or dummy nodes, provided the most optimal performance. Multi-fidelity transfer may also improve performance \citep{Thiele:2025mlo}.


Our study focused on a simple fully connected network, but we expect qualitatively similar conclusions for more expressive architectures such as normalizing flows predicting the full posterior distribution, which would be a natural extension to test. Moreover, while we restricted our analysis to matter power spectra, applying transfer learning to observables such as galaxy clustering or weak lensing would be fruitful future work---in some cases this may yield greater gains, as, for example in the neutrino mass case, these observables have reduced sensitivity to $M_\nu$ \citep{Bayer_2022_fake}, making an easier transfer task. Looking beyond cosmology, this analysis could inform other areas of fundamental physics, such as learning extensions beyond the Standard Model of particle physics.

Overall, our results suggest that transfer learning can accelerate inference beyond the standard model, but its effectiveness hinges on parameter degeneracies, the choice of data summary, and the choice of architecture. More broadly, they illustrate both the promise and the pitfalls of foundation models for physics: pre-training on large standard-model datasets can dramatically reduce costs, but may also bias representations in ways that hinder the discovery of new physics if not carefully safeguarded.

\bibliographystyle{plainnat}
\bibliography{references}

\appendix

\section{Data Description}
\label{app:data}

Here we provide a thorough description of the Quijote simulations\footnote{\url{https://quijote-simulations.readthedocs.io/}} \citep{Quijote_sims}. For the $\Lambda$CDM (pre-training) simulation, we use the Big Sobol Sequence (BSQ) suite. BSQ consists of 32,768 simulations described by 5 cosmological parameters with varying values: $\Omega_m \in [0.10,\ 0.50],\ \Omega_b \in [0.02,\ 0.08],\ h \in [0.50,\ 0.90],\ n_s \in [0.80,\ 1.20], \ \sigma_8 \in [0.60,\ 1.00]$. In all simulations $M_{\nu} = 0{\rm eV}$, $w = -1$, $f_{R0}=$, and $f_{\rm NL}=0$. 

The three different beyond-$\Lambda$CDM (fine-tuning) simulations setups are described as follows:
\begin{itemize}
    \item For $M_\nu$ we use 2,000 Latin Hypercube simulations which vary $M_{\nu}$ in the range $M_{\nu} \in [0.01,\ 1.0]$eV and $w$ in the range $w \in [-1.3,\ -0.7]$. While $w$ is varied, it cannot be constrained with a single redshift snapshot in real space, as it only affects the cosmological background, so we do not perform inference on $w$ here. Initial tests confirmed that the network failed to learn any meaningful information about $w$, and including it in the inference task only appeared to weaken performance. 
    The ranges of the five $\Lambda$CDM parameters match those of the BSQ, with the exception of $\Omega_b \in [0.03,\ 0.07]$. We consider two summary statistics in this case, the power spectrum $P$ and the marked power spectrum $MP$, as it has been shown that $MP$ is much more constraining on $M_\nu$  compared to $P$ \cite{Massara:2020pli}: this enables comparison in transfer learning on the amount of information in the summary.
    \item In the case of modified gravity, the Quijote simulations use a Hu and Sawicki $f(R)$ model \cite{Hu_and_Sawicki} where the Einstein-Hilbert action is extended by a function of the Ricci scalar, introducing a scalar degree of freedom that modifies gravity on large scales. For $f(R)$, we use 2,048 simulations with the $\Lambda \rm{CDM}$ parameters following the same ranges as in BSQ, with the addition of $M_{\nu} \in [0.01,\ 1.0]$eV and $f_{\text{R0}} \in [-3 \times 10^{-4},\ 0]$. With modified gravity there are two definitions of $\sigma_8$: one corresponding to the GR underlying cosmology ($\sigma_8$(LCDM)) and another reflecting the full modified gravity model ($\sigma_8$(MG)). For the purpose of transfer learning, we perform inference on $\sigma_8$(LCDM) for consistency.
    \item For $f_{\rm NL}$ we consider local ($f_{\text{NL}}^{\text{local}}$) and equilateral ($f_{\text{NL}}^{\text{equilateral}}$) using the Quijote-PNG suite \cite{PNG}. For each, we use a Latin hypercube with 1,000 simulations. The local set fixes $\Omega_m = 0.3175$, $\Omega_b = 0.049$, $h = 0.6711$, $n_s = 0.9624$, $\sigma_8 = 0.834$, and $M_{\nu} = 0\ \text{eV}$, while varying $f_{\mathrm{NL}}^{\mathrm{local}} \in [-300,\ 300]$. The equilateral set keeps $\Omega_b = 0.049$ and $M_{\nu} = 0$eV fixed, and varies the remaining $\Lambda$CDM parameters as in BSQ, along with $f_{\text{NL}}^{\text{equilateral}} \in [-600,\ 600]$. This allows us to test the effects of transfer learning when the prior on $\Lambda \rm{CDM}$ parameters differs between the two sets. 
\end{itemize}

All simulations follow the evolution of $512^3$ dark matter particles in a periodic comoving volume of $(1\ h^{-1} \text{Gpc})^3$, with initial conditions generated at $z = 127$ and evolved using Gadget-III. Simulations that include massive neutrinos add an additional $512^3$ neutrino particles. 
Although the parameter ranges are mostly consistent across cosmologies, $\Omega_b$ varies slightly; for normalization consistency across models, we adopt the broader range defined by the $\Lambda$CDM dataset for normalizations.

\section{Additional Results}

\subsection{MSE for all parameters}

Here we provide results and discussion of the MSE on all the cosmological parameters for all the different examples considered in the paper. We also test further options for the number of pre-training simulations.

\begin{figure}[h] 
    \centering
    \begin{subfigure}{0.32\linewidth}
        \centering
        \includegraphics[width=\linewidth]{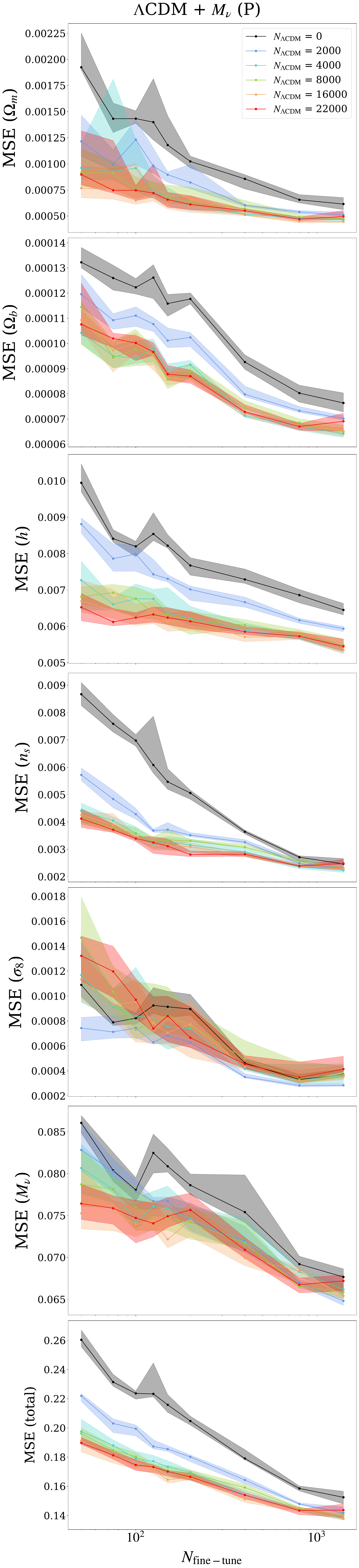}
        \caption{}
        \label{fig:mse_nwLH_Pk_full}
    \end{subfigure}
    \hspace{0.0\linewidth}
    \begin{subfigure}{0.32\linewidth}
        \centering
        \includegraphics[width=\linewidth]{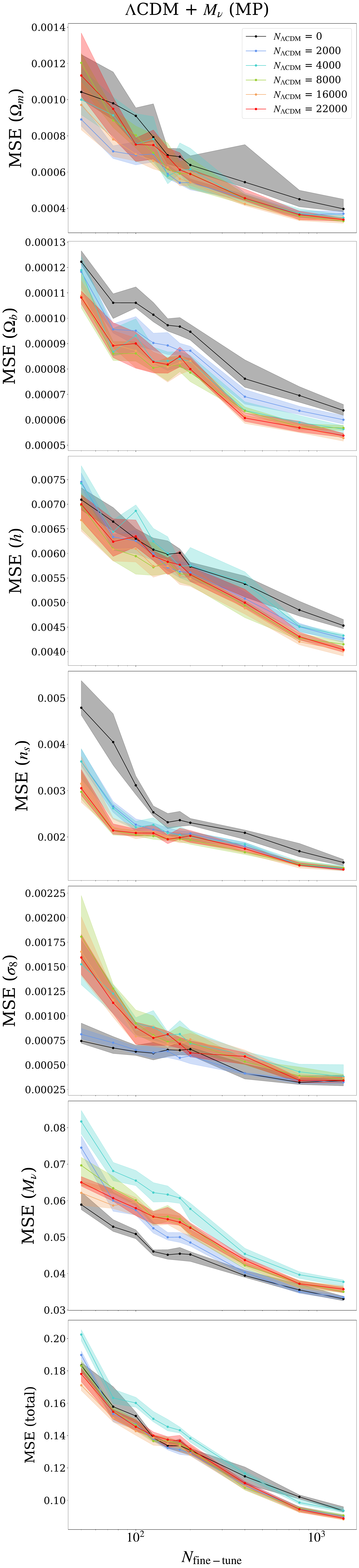}
        \caption{}
        \label{fig:mse_nwLH_MPk_full}
    \end{subfigure}
    \caption{Extension of Figure~\ref{fig:mse_nwLH} showing the MSE for all individual parameters in the massive neutrino cosmology, but only for the dummy node architecture. Transfer learning provides improvements for some $\Lambda$CDM parameters when training data is very limited and when using the power spectrum (left), but offers little to no benefit for $\sigma_8$ and $M_\nu$. In fact, for the marked power spectrum (right) it can even degrade performance (negative transfer) at low simulation counts.
}
    \label{fig:mse_nwLH_full}
\end{figure}

For massive neutrinos with the standard power spectrum (Figure~\ref{fig:mse_nwLH_Pk_full}), transfer learning modestly improves performance for some $\Lambda$CDM parameters when training data is exceptionally scarce. However, it offers little to no benefit for $\sigma_8$ and $M_{\nu}$, even at low simulation counts. At larger training set sizes, training both with and without transfer learning yields similar results.

A similar trend is observed when using the marked power spectrum (Figure~\ref{fig:mse_nwLH_MPk_full}), where all parameters show either some improvements or comparable performance when using transfer learning. However, unlike the standard power spectrum, transfer learning does not offer a significant advantage at any number of training simulations. Furthermore, at low numbers of beyond-$\Lambda$CDM simulations, transfer learning performs noticeably worse than training from scratch for $M_{\nu}$ and $\sigma_8$ in particular. This decline is likely driven by degeneracy between $M_\nu$, $\sigma_8$, and $\Omega_m$. During pre-training, the marked power spectrum learns a precise knowledge of $\sigma_8$ and $\Omega_m$ which then has to be unlearned in order to recognize the effects of $M_\nu$ when it is introduced. This is an example of \textit{negative transfer}. We do not observe the same behavior for the power spectrum because it is not informative of $M_\nu$ on its own \cite{Bayer_2022_fake} and thus the introduction of $M_\nu>0$ does not confuse the pre-trained network. 

\begin{figure}[h] 
    \centering
    \begin{subfigure}{0.3\linewidth}
        \centering
        \includegraphics[width=\linewidth]{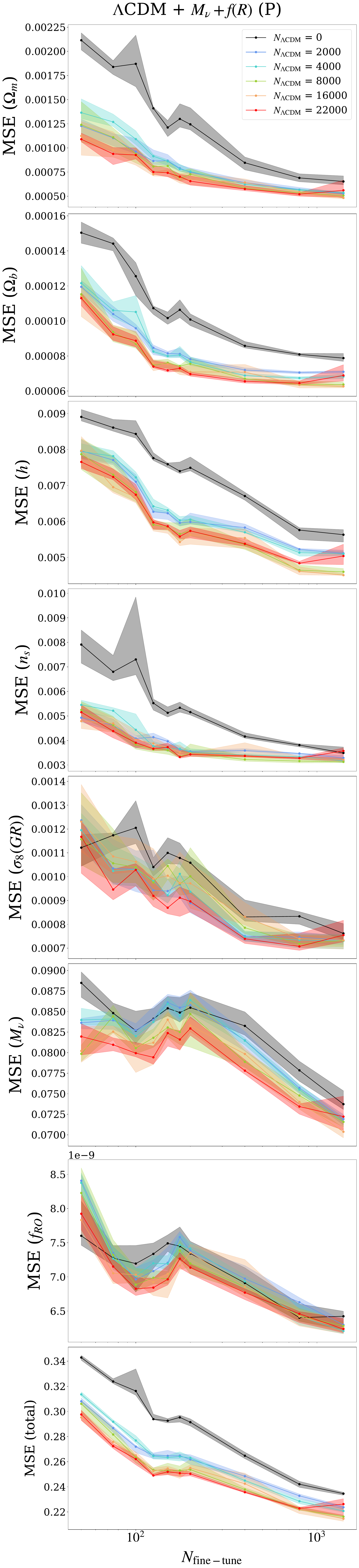}
        \caption{}
        \label{fig:mse_fR_Pk_full}    
    \end{subfigure}
    \hspace{0.0\linewidth}
    \begin{subfigure}{0.3\linewidth}
        \centering
        \includegraphics[width=\linewidth]{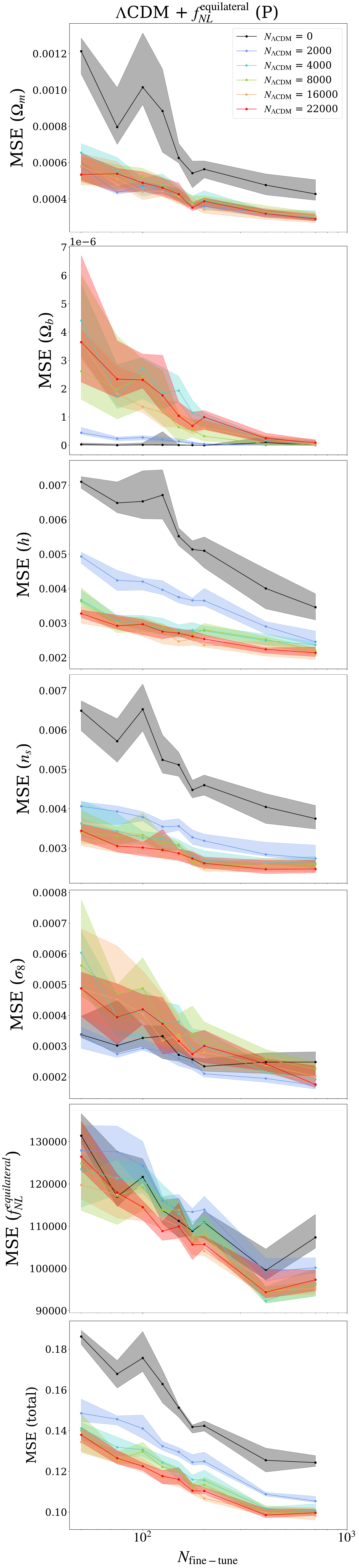}
        \caption{}
        \label{fig:mse_EQ_Pk_full}
    \end{subfigure}
    \hspace{0.0\linewidth}
    \begin{subfigure}{0.3\linewidth}
        \centering
        \includegraphics[width=\linewidth]{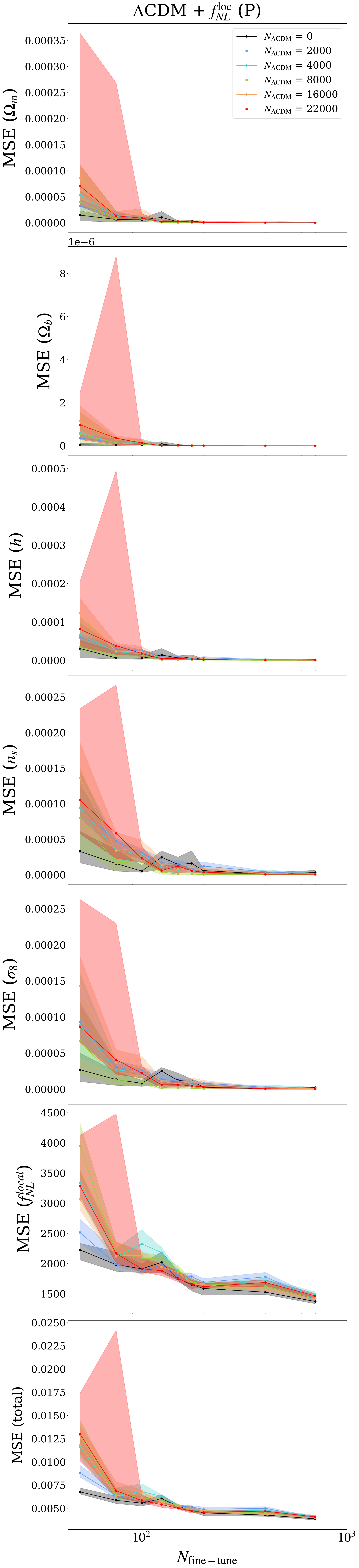}
        \caption{}
        \label{fig:mse_LC_Pk_full}
    \end{subfigure}
    \caption{Same as Figure~\ref{fig:mse_beyond_lcdm}, but showing MSE for each individual parameter in the modified gravity, equilateral, and local non-Gaussianity cosmologies.}
    \label{fig:mse_beyond_lcdm_full}
\end{figure}

We now consider the combination of massive neutrinos and modified gravity in Figure~\ref{fig:mse_fR_Pk_full}. Transfer learning provides a noticeable advantage, particularly when training data is limited, mirroring trends seen in \ref{fig:mse_nwLH_Pk_full}. For $f_R{_0}$, $M_\nu$ and $\sigma_8$, performance is nearly identical with and without transfer learning -- consistent with earlier power spectrum results for massive neutrinos, and again likely due to degeneracies between these parameters. However, unlike the marked power spectrum case, no significant performance drop is observed here, reinforcing the idea that the decline there stems from the marked power spectrum’s added sensitivity to $M_\nu$.

For equilateral-type primordial non-Gaussianity (Figure~\ref{fig:mse_EQ_Pk_full}), where all parameters except $\Omega_b$ are varied, transfer learning consistently outperforms the baseline for all parameters except $\sigma_8$ and $\Omega_b$, where performance is similar. This suggests that while degeneracy between $\sigma_8$ and $f_{\text{NL}}^{\text{equilateral}}$ may limit gains for those parameters, the influence of $f_{\text{NL}}$ on the other cosmological parameters is minimal, allowing transfer learning to provide positive results in those cases.

For the local primordial non-Gaussianity case (Figure~\ref{fig:mse_LC_Pk_full}) -- where all $\Lambda$CDM parameters are fixed across the fine-tuning data -- transfer learning offers no benefit over direct training on the beyond-$\Lambda$CDM dataset, with nearly identical results across all parameters. Since the fine-tuning task only involves learning the effect of $f_{\rm NL}^{\rm local}$ on the power spectrum, transfer learning appears unnecessary and ineffective, but does also not hinder performance. 

\label{app:additional_results}

\subsection{Alternative architectures}
\label{app:arch}

We also tested two alternative pre-training setups:
\begin{enumerate}
    \item No-dummy pre-training:  
    This network is identical to the setup described in Section~\ref{sec:methods}, except that no dummy output node was included during pre-training. In this case, for fine-tuning only the $\Lambda$CDM parameters were initialized with pre-trained weights, while the additional parameters required for the beyond-$\Lambda$CDM models started from random initialization.  

    \item Attach a trainable inference head:  
    Here we modified the pre-training network from Section~\ref{sec:methods} by constraining the final hidden layer to 10 neurons (finding this to be optimal). Once trained, we passed power spectra from the smaller beyond-$\Lambda$CDM datasets (those including massive neutrinos, modified gravity, or primordial non-Gaussianities) through the best performing $\Lambda$CDM model and extract the 10-dimensional output of the final hidden layer. 
    This effectively reduces each beyond-$\Lambda$CDM power spectrum to a set of 10 latent features. In the second stage we train a new network on these reduced power spectra to predict the extended set of cosmological parameters of each beyond-$\Lambda$CDM cosmology.
\end{enumerate}

\begin{figure}[h] 
    \centering
    \begin{subfigure}{0.32\linewidth}
        \centering
        \includegraphics[width=\linewidth]{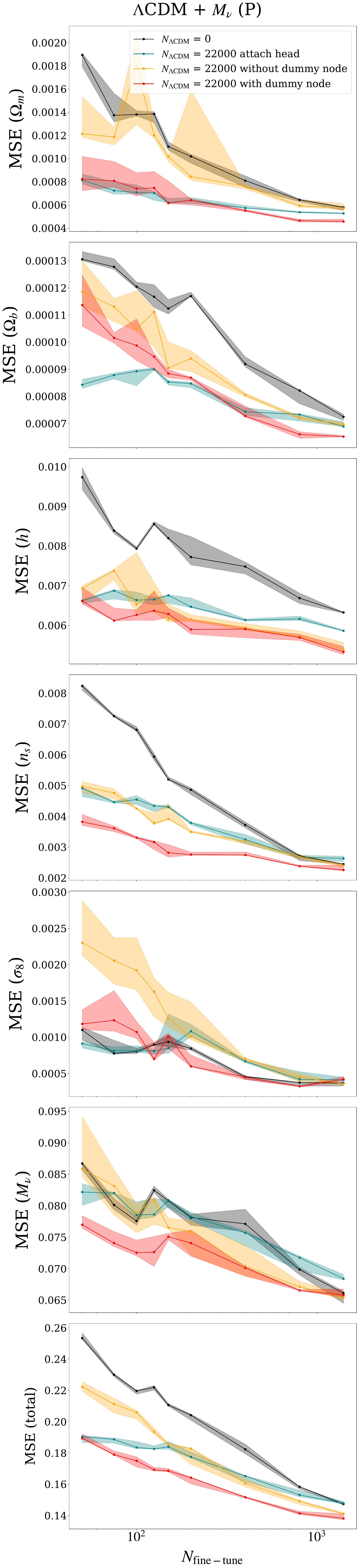}
    \end{subfigure}
    \hspace{0.0\linewidth}
    \begin{subfigure}{0.32\linewidth}
        \centering
        \includegraphics[width=\linewidth]{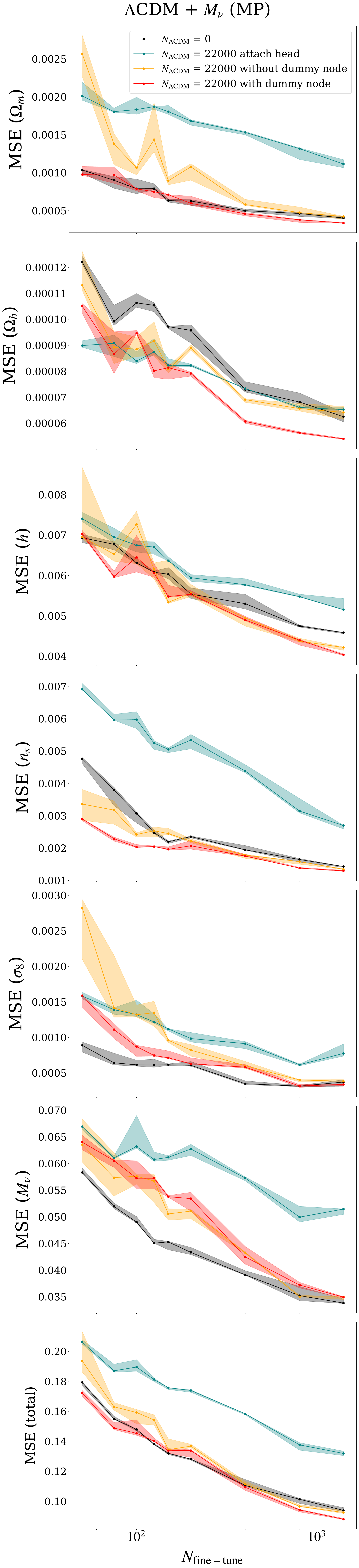}
    \end{subfigure}
    \caption{Same as Figure~\ref{fig:mse_nwLH}, but showing the MSE for all individual parameters in the massive neutrino cosmology. Provides full context for per-parameter behavior discussed in the main text in Section~\ref{sec:results}.}
    \label{fig:mse_nwLH_ft_test_full}
\end{figure}

Overall, we found that the latent-feature head case performed the worst, often exhibiting severe negative transfer. The weight-initialization case performed better, but the dummy-node setup described in Section~\ref{sec:methods} was the most effective, as shown in Fig.~\ref{fig:mse_nwLH_ft_test_full}.

\subsection{Parameter Degeneracy Analysis}
\label{app:shap}

 \begin{figure}[h] 
 \vspace{-1.5em}
     \centering
     \hspace{-4em}
     \begin{subfigure}{0.35\linewidth}
         \centering
         \includegraphics[width=\linewidth, height=18cm]{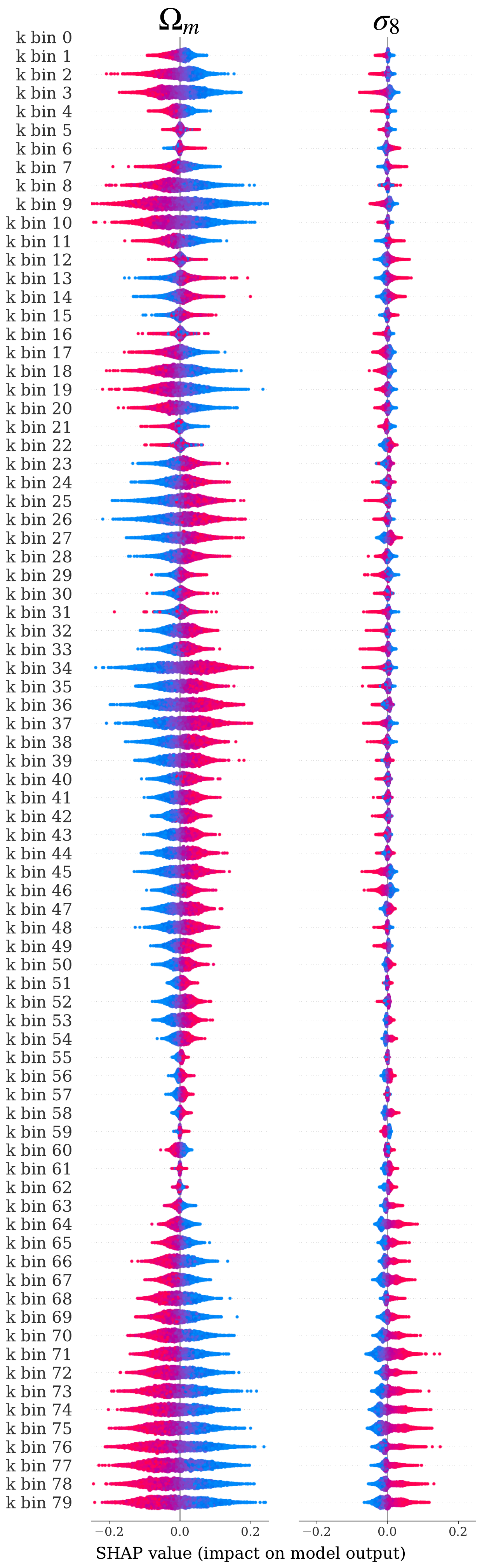}
         \caption{}
         \label{fig:SHAP_pretrain}
     \end{subfigure}
     \hspace{0.05\linewidth}
     \begin{subfigure}{0.25\linewidth}
         \centering
         \includegraphics[height=18cm]{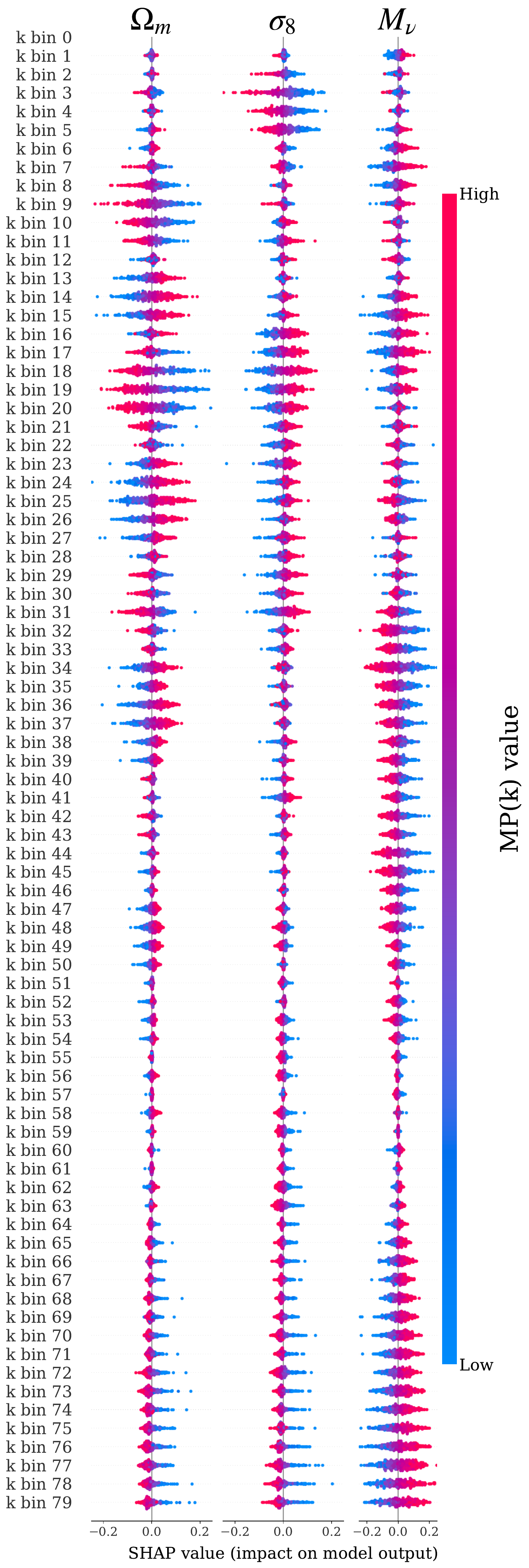}
         \caption{}
         \label{fig:SHAP_finetune}
     \end{subfigure}
     \caption{\texttt{SHAP beeswarm} plots for $\Omega_m$ and $\sigma_8$ in the pretrained model (left) and for $\Omega_m$, $\sigma_8$, and $M_\nu$ in the fine-tuned model (right), computed on the marked power spectrum MP$(k)$. SHAP values quantify the local contribution of a feature ($y$-axis) to the model output relative to a baseline. The sign of the SHAP value ($x$-axis) indicates whether increasing that feature pushes the prediction up or down and the horizontal spread at a given feature indicates importance. Here we consider the power-spectrum $k$ bins as the features, while the color represents the value of $P(k)$. In pretraining, small scales i.e. high-$k$ bins carry substantial contribution for $\sigma_8$. After introducing $M_\nu$, that small-scale influence is reassigned to $M_\nu$, while $\sigma_8$ shows a sign flip at high $k$ (i.e high power (pink) pushed the $\sigma_8$ prediction up (positive SHAP) and low power (blue) pushed the prediction down (negative SHAP), but during fine-tuning $M_\nu$ adopts this behavior and $\sigma_8$'s is reversed) and its relative weight shifts toward larger scales. This pattern indicates that the model’s initial small-scale $\sigma_8$ cue is “unlearned” and repurposed for $M_\nu$ — indicative of the $\sigma_8$–$M_\nu$ degeneracy that underlies the observed negative transfer.
 }
     \label{fig:SHAP}
 \end{figure}

To assess how degeneracies among $\Omega_m$, $\sigma_8$, and $M_{\nu}$ may shape the network's performance, we study which parts of the marked power spectrum the model relies on to infer each parameter, as shown in Figure~\ref{fig:SHAP}. We compare \texttt{SHAP beeswarm} plots for a pretrained model trained on $22{,}000$ $\Lambda$CDM simulations (Figure~\ref{fig:SHAP_pretrain}) and a fine-tuned model trained on 50 massive-neutrino simulations using the marked power spectrum (Figure~\ref{fig:SHAP_finetune}), where negative transfer is most pronounced at low beyond-$\Lambda$CDM sample counts.

During pretraining, the network learns to attribute small-scale power-spectrum variations to $\Omega_m$ and $\sigma_8$, forming a LCDM-consistent mapping of those features to parameters. When $M_\nu$, which physically suppresses small-scale power, is introduced, those same modes become predictive for $M_\nu$, forcing the model to reassign small-scale sensitivity. This reallocation is consistent with the model treating $M_\nu$-driven changes as if they were $\sigma_8$-like under the pretrained representation. Under fine-tuning $\sigma_8$ is still important at small scales but with a reversed sign and has a grater reliance on large-scale information. By contrast, while $\Omega_m$ loses some small-scale influence, its oscillatory sensitivity pattern at large scales remains comparatively unchanged, indicating more transferable structure. Overall, it appears that during fine-tuning the network has to effectively ``unlearn'' its $\sigma_8$ mapping at small scales and reallocate it to a combination of $M_\nu$ and $\sigma_8$ effects and ``relearn'' $\sigma_8$ from elsewhere, resulting in degradation in performance and negative transfer.

\end{document}